\documentclass[sii]{ipart}

\RequirePackage[OT1]{fontenc}
\RequirePackage{amsthm,amsmath}
\RequirePackage[numbers,square]{natbib}
\RequirePackage[colorlinks]{hyperref}
\usepackage{algorithm}
\usepackage{algpseudocode}
\usepackage{graphicx}
\usepackage{booktabs}

\newtheorem{theorem}{Theorem}

\pubyear{2014}
\volume{0}
\issue{0}
\firstpage{1}
\lastpage{1}

\startlocaldefs
\endlocaldefs

\begin{document}

\begin{frontmatter}

\title{Nonparametric Statistical Inference and Imputation for Incomplete Categorical Data\thanksref{T1}}
\thankstext{T1}{The authors gratefully acknowledge three grants from the Research Grants Council of the Hong Kong SAR, China (400913, 14203915, 14173817) and two CUHK direct grants (4053135, 3132753).}
\runtitle{Inference and Imputation for Incomplete Categorical Data}

\begin{aug}
\author{\fnms{Chaojie}
\snm{Wang}\ead[label=e1]{wang910930@gmail.com}},
\address{Faculty of Science, Jiangsu University\\
Zhenjiang, China\\
Department of Statistics, The Chinese University of Hong Kong\\
Shatin, N.T., Hong Kong\\
\printead{e1}}
\author{\fnms{Linghao} \snm{Shen}\ead[label=e2]{markshen1991@163.com}},
\address{Department of Information Engineering\\
The Chinese University of Hong Kong\\
Shatin, N.T., Hong Kong\\
\printead{e2}}
\author{\fnms{Han} \snm{Li}
\ead[label=e3]{hli@szu.edu.cn}}
\address{Department of Risk Management and Insurance\\
Shenzhen University\\
Shenzhen, China\\
\printead{e3}}
\and
\author{\fnms{Xiaodan} \snm{Fan}\thanksref{t1}\corref{}
\ead[label=e4]{xfan@cuhk.edu.hk}}
\address{Department of Statistics, The Chinese University of Hong Kong\\
Shatin, N.T., Hong Kong\\
\printead{e4}}
\thankstext{t1}{Corresponding author: xfan@cuhk.edu.hk}

\runauthor{Wang et al.}

\affiliation{Jiangsu University and The Chinese University of Hong Kong}
\end{aug}

\begin{abstract}
Missingness in categorical data is a common problem in various real applications. Traditional approaches either utilize only the complete observations or impute the missing data by some ad hoc methods rather than the true conditional distribution of the missing data, thus losing or distorting the rich information in the partial observations. In this paper, we propose the Dirichlet Process Mixture of Collapsed Product-Multinomials (DPMCPM) to model the full data jointly and compute the model efficiently. By fitting an infinite mixture of product-multinomial distributions,
  DPMCPM is applicable for any categorical data regardless of the true distribution, which may contain
  complex association among variables. Under the framework of latent class analysis, we show that DPMCPM
  can model general missing mechanisms by creating an extra category to denote missingness, which implicitly integrates out the missing part with regard to their true conditional distribution.
  Through simulation studies and a real application, we demonstrate that DPMCPM outperforms
  existing approaches on statistical inference and imputation for incomplete categorical data of various missing mechanisms. DPMCPM is implemented as the R package \texttt{MMDai}, which is available from the Comprehensive R
Archive Network at https://cran.r-project.org/web/packages/MMDai/index.html.
\end{abstract}





\end{frontmatter}

\section{Introduction}
\label{sec:introduction}

Missingness in categorical data is a common problem in many real
applications. For examples, in social surveys, data collected by
questionnaires are often incomplete because subjects may be
unwilling or unable to respond to some items
\cite{kuha2018latent}. In biological experiments,
data may be
incomplete for either biologically-driven or technically-driven reasons
\cite{hicks2017missing}. In recommendation system problems
\cite{ricci2015recommender}, analysts often have a dataset as the
toy example in Table~\ref{table1}. The goal is to predict the
potential purchase behavior without direct observation, e.g.,
whether Customer3 and Customer4 would buy Book5.

\begin{table}[!htbp]
\centering
\begin{tabular}{|c|c|c|c|c|c|}
  \hline
   & Book1 & Book2 & Book3 & Book4 & Book5 \\
  \hline
  Customer1 & 1 & 1 &  &  & 0 \\
  \hline
  Customer2 & 1 & 0 &  &  & 0 \\
  \hline
  Customer3 &  & 1 & 1 &  &  \\
  \hline
  Customer4 &  & 1 & 1 &  &  \\
  \hline
  Customer5 &  &  & 0 & 1 & 0 \\
  \hline
  Customer6 &  &  & 0 & 1 & 0 \\
  \hline
\end{tabular}
\caption{Toy data example with the goal to infer the empty cells. 1: bought; 0: recommended but not
bought; empty: missing values.}\label{table1}
\end{table}

When dealing with datasets with missingness, naive approaches,
such as the complete-case analysis (CCA) and overall mean imputation,
would waste the information in the missing data and may bias
the inference \cite{donders2006review}. When missingness is high, CCA is hardly applicable due to the lack of complete cases.
Advanced methods, such as multiple imputation \cite{rubin1987multiple,schafer2002missing}, impose a parametric model on the data and then draw multiple sets of samples to account for the uncertainty of
the missing information. For categorical cases,  \cite{schafer1997analysis}
advocated the log-linear model for multiple imputation, which can
capture certain types of association among the categorical
variables. However, this model works only when the number of
variables is small, as the full multi-way cross-tabulation
required for the log-linear analysis increases exponentially with
the number of variables \cite{vermunt2008multiple}.

There are two basic ideas for imputing multivariate missing data: fully conditional specification (FCS) and joint modeling. FSC \cite{van2006fully} specifies a collection of univariate conditional imputation models that condition on all the other variables. A popular application based on FCS is known as Multiple Imputation by Chained
Equation (MICE), which specifies a sequence of regression models
iteratively \cite{van1999flexible,white2011multiple}.  \cite{buuren2011mice}
provided the R package \texttt{mice} to implement this
method efficiently. Although it has been shown to work well for many datasets
and become a popular method \cite{jolani2015imputation}, MICE still has some common drawbacks of FCS. A typical application of MICE is to use multinomial logistic regression for the categorical data, but the relationship among
variables may be nonlinear and may involve complex interaction or
higher-order effects \cite{murray2016multiple}. Besides,
there is no guarantee that the iterations of sequential regression
model will converge to the true posterior distribution of missing
values \cite{vermunt2008multiple}. Other parametric approaches include missMDA based on principal component analysis \cite{josse2016missmda}, MIMCA based on correspondence analysis \cite{audigier2017mimca}, and so on. These parametric methods can be applied in some specific problems but not in general cases. \cite{murray2018multiple} provided a detailed review for current advances of multiple imputation.

From the perspective of data analysts, no matter the data is raw with missingness or has been imputed when arriving, it is important to understand the detailed mechanisms of pre-processing, including imputation. \cite{xie2017dissecting} reviewed and discussed the imputation danger caused by the difference of the God's model, imputer's model and analyst's model. They pointed that any attempt of pre-processing to make the data ``more usable'' implies potential assumptions. In this aspect, joint modeling provides a powerful tool to model the underlying distribution behind the data. For categorical data,
\cite{vermunt2008multiple} proposed a latent class model, i.e.,
the finite mixture of product-multinomial model.
The latent class model can characterize both simple association
and complex higher-order interactions among the categorical
variables as long as the number of latent class is large enough
\cite{mclachlan2000finite}. \cite{dunson2009nonparametric}
proposed the Dirichlet Process Mixture of Products of Multinomial distributions (DPMPM) to model \emph{complete} multivariate categorical datasets, which
avoids the arbitrary setting of the number of mixture components.
Furthermore, \cite{dunson2009nonparametric} proved that any
multivariate categorical data distribution can be approximated by
DPMPM for a sufficiently large number of mixture components.
\cite{si2013nonparametric} generalized the DPMPM framework to analyze incomplete categorical datasets, which works well for low missingness but performs poorly for high missingness since the number of parameters would increase dramatically. Based on the work of \cite{si2013nonparametric}, \cite{manrique2017bayesian} introduced a variant of this model for edit-imputation, which accounts for the values that are logically impossible but present due to measurement error. \cite{hu2018dirichlet} extended this model to nested data structures in the presence of structural zeros. Other related works include the divisive latent class model \cite{van2015divisive}, Bayesian multilevel latent class
models \cite{vidotto2018bayesian} and so on.
\cite{vidotto2014multiple} presented a detailed overview of recent researches on multiple imputation using the latent class model.

In this paper, we propose DPMCPM, which extends DPMPM for
modelling \emph{incomplete} multivariate categorical data efficiently.
DPMCPM inherits some nice properties of DPMPM. It avoids the
arbitrary setting of the number of mixture components by using the
Dirichlet process. In addition, DPMCPM is applicable for any
categorical data regardless of the true distribution and can
capture the complex association among variables.
Different from the missingness viewed as unknown parameters in \cite{si2013nonparametric}, DPMCPM
creates an extra category to denote the
missingness. It shall reduce computation burden and gain better performance when the missingness is high. Under the framework of the latent class analysis, we show that DPMCPM can model
general missing mechanisms.
Through simulation
studies and a real application, we demonstrate DPMCPM performs better statistical
inference and imputation than existing approaches. To our
knowledge, this is the first non-parametric tool which can model
arbitrary categorical distributions and handle high missingness.

This paper is organized as follows. Section~\ref{sec:method}
introduces the DPMCPM model and the Gibbs sampler algorithm.
Section~\ref{sec:sim} performs simulation studies on the synthetic
data and Section~\ref{sec:app} presents a real application in a
recommendation system problem. Section~\ref{sec:summary} concludes
the paper.

\section{Method}
\label{sec:method}

\subsection{Dirichlet Process Mixture of Product-Multinomials}\label{sec:DPMPM}
We begin with the finite mixture product-multinomial model for the
case of complete dataset $\boldsymbol{x}=\{x_{ij}\}$, where
$i=1,\cdots,n$ and $j=1,\cdots,p$. Suppose $\boldsymbol{x}$
comprises of $n$ independent samples associated with $p$
categorical variables, and the $j$-th variable has $d_j$
categories. Let $x_{ij}\in\{1,\cdots,d_j\}$ denote the observed
category for the $i$-th sample in the $j$-th variable. The finite
mixture product-multinomial model assumes that those $x_{ij}$ are
generated from a multinomial distribution indexed by a latent
variable $z_i\in\{1,\cdots,k\}$. A finite mixture model with $k$
latent components can be expressed as: \begin{equation}
\label{eq1}
x_{ij}|z_i,\boldsymbol{\psi}_{z_i}^{(j)}\sim multinomial(\psi_{z_i1}^{(j)},\cdots,\psi_{z_id_j}^{(j)}),\\
\end{equation}
\begin{equation}
\label{eq2}
z_i|\Theta\sim multinomial(\theta_1,\cdots,\theta_k),\\
\end{equation}
where $\Theta=\{\theta_1,\cdots,\theta_k\}$ and $\boldsymbol{\psi}_{z_i}^{(j)}=\{\psi_{z_i1}^{(j)},\cdots,\psi_{z_id_j}^{(j)}\}$. We further define $\Psi=\{\boldsymbol{\psi}_{h}^{(j)}:h=1,\cdots,k;j=1,\cdots,p\}$ and $\boldsymbol{z}=\{z_{i}:i=1,\cdots,n\}$.

\cite{dunson2009nonparametric} proved that any multivariate
categorical data distribution can be represented by the mixture
distribution in Equation~\ref{eq1} and \ref{eq2} for a
sufficiently large $k$. However, specifying a good $k$ to avoid
over-fitting and over-simplification is non-trivial, and it
becomes even harder when the dataset is highly incomplete
\cite{dunson2009nonparametric,si2013nonparametric}. This motivates
the use of an infinite extension of the finite mixture model,
i.e., the Dirichlet process mixture. A Dirichlet process can be
represented by various schemes, including the P\'{o}lya urn
scheme, the Chinese restaurant process and the stick-breaking
construction \cite{teh2006hierarchical}.
\cite{dunson2009nonparametric} chose the stick-breaking
construction to model the Dirichlet process. However, in practice,
the slice Gibbs sampler in their construction may often be trapped
in a single component when $n$ is relatively large due to numeric
limits and thus fail to identify the correct number of components \cite{walker2007sampling}. To avoid this drawback, we
construct the Dirichlet process by using the Chinese restaurant
process in DPMCPM:
\begin{equation}
\begin{split}
\label{eq3}
&P(z_i=h|\boldsymbol{z}_{-i})=\frac{n_{h,-i}}{n+\alpha-1},~~h=1,\cdots,k;\\
&P(z_i=k+1|\boldsymbol{z}_{-i})=\frac{\alpha}{n+\alpha-1},\\
\end{split}
\end{equation}
where $k$ denotes the number of previously occupied components and
$n_{h,-i}$ denotes the number of samples in the $h$-th component
excluding the $i$-th sample. Here $\alpha$ is a prior
hyper-parameter that acts as the pseudo-count of the number of
samples in the new component regardless of the input data.

\subsection{Multinomials with Incomplete Data}\label{sec:incomplete}
\cite{rubin1976inference}
introduced a class of missing mechanisms, which are commonly
referred as Missing Completely At Random (MCAR), Missing At Random
(MAR) and Missing Not At Random (MNAR). Traditionally, missing mechanisms of a data matrix $\boldsymbol{x}$ are denoted by introducing an indicator matrix $\boldsymbol{r}=(r_{ij})$, where $r_{ij}=1$ if $x_{ij}$ is observed, and $r_{ij}=0$ if $x_{ij}$ is missing. The missing rates $p(r_{ij}|\boldsymbol{x}_i)$ under three types of missing mechanisms are defined as follows:
\begin{itemize}
 \item MCAR: Missingness does not depend on the missing or observed data, i.e.,
 \begin{equation*}
 p(r_{ij}|\boldsymbol{x}_i)=\tau,
 \end{equation*} where $\boldsymbol{x}_i=(x_{i1},\cdots,x_{ip})$ is the $i$-th observation and $\tau\in(0,1)$ is a constant which does not depend on $\boldsymbol{x_i}$;

 \item MAR: Missingness only depends on the observed data, i.e.,
     \begin{equation*}
 p(r_{ij}|\boldsymbol{x}_i)=p(r_{ij}|\boldsymbol{x}_{i,obs}),
 \end{equation*}
     where $\boldsymbol{x}_{i,obs}$ denotes the observed values in the $i$-th observation;

 \item MNAR: Missingness depends
on both the missing and observed data, i.e.,
     \begin{equation*}
 p(r_{ij}|\boldsymbol{x}_i)=p(r_{ij}|\boldsymbol{x}_{i,obs},\boldsymbol{x}_{i,mis}),
 \end{equation*}
     where $\boldsymbol{x}_{i,obs}$ and $\boldsymbol{x}_{i,mis}$ denote the observed and missing values in the $i$-th observation, respectively.
\end{itemize}

In DPMCPM, we model the incomplete data by creating an extra
category to denote the missingness. Here Equation~\ref{eq4} is
used to replace Equation~\ref{eq1} in the DPMPM model:
\begin{equation}
\label{eq4}
x_{ij}|z_i,\boldsymbol{\psi}_{z_i}^{(j)}\sim multinomial(\psi_{z_i0}^{(j)},\psi_{z_i1}^{(j)},\cdots,\psi_{z_id_j}^{(j)}),\\
\end{equation}
where the extra $0$-th category denotes the case when $x_{ij}$ is
missing.

Once we get the parameter estimates using Equation~\ref{eq4}, we
shall rescale the multinomial probabilities as follows to recover
the corresponding parameters in Equation~\ref{eq1}:
\begin{equation*}
\tilde{\psi}_{hc}^{(j)}=p(x_{ij}=c|x_{ij}\neq 0,z_i=h,\boldsymbol{\psi}_{h}^{(j)})={\psi_{hc}^{(j)}}/{(1-\psi_{h0}^{(j)})},
\end{equation*}
for $c=1,\cdots,d_j$. This procedure implicitly integrates out the missing values
according to their estimated conditional distributions. As compared
to \cite{si2013nonparametric}, this ``collapse'' step dramatically
shrinks the dimension of the posterior sampling space, thus
enables DPMCPM to handle high missingness.

The idea of introducing an extra category comes from \cite{formann2007mixture} for the
log-linear model but they did not provide a strict proof. In fact, because any relationship among
variables can be approximated by the mixture product-multinomial
model for a sufficiently large $k$, DPMCPM also accommodates any
dependencies among the $0$-th categories and other parameters,
thus it can capture the MCAR, MAR and MNAR missing mechanisms
\cite{dunson2009nonparametric,formann2007mixture}. We have the following theorem:
\begin{theorem}
For any general missing rate $p(r_{ij}|\boldsymbol{x}_{i})$, there exists at least one set of parameters $\boldsymbol{\Theta}$ and
$\boldsymbol{\Psi}=\{{\psi}_{hc}^{(j)}:h=1,\cdots,k;j=1,\cdots,p;c=0,1,\cdots,d_j\}$ in DPMCPM, which is equivalent with the corresponding missing rate.
\end{theorem}

\emph{Proof:} See Appendix \ref{appA} for detailed proofs.

Based on the estimation of $\boldsymbol{\Theta}$ and
$\tilde{\boldsymbol{\Psi}}=\{\tilde{\psi}_{hc}^{(j)}:h=1,\cdots,k;j=1,\cdots,p;c=1,\cdots,d_j\}$, we can obtain an approximate
of the true distribution up to the Monte Carlo error.
With this estimated distribution, any statistical inference and
imputation can be performed. For example, if $x_{ij}$ is missing
and a single imputation is desired, we can impute $\hat x_{ij}=c_{\text{pred}}$
by
\begin{equation*}
c_{\text{pred}}=\arg\max_cp(x_{ij}=c|\boldsymbol{x}_{i,obs},\boldsymbol{\Theta},\tilde{\boldsymbol{\Psi}}),
\end{equation*}
where $\boldsymbol{x}_{i,obs}$ denotes all of the observed data in
the $i$-th sample and $c_{\text{pred}}$ is the predicted value.

\subsection{Posterior Inference}\label{sec:posterior}
We use the Bayesian approach to fit the non-parametric model in
Equation~\ref{eq2}-\ref{eq4} to the data. More specifically, we
introduce prior distributions for unknown parameters, and then use
a Markov chain Monte Carlo algorithm to sample from their posterior
distribution. The converged samples will be used for statistical
inference.

For prior distributions, we assume the conjugate prior for
$\boldsymbol{\psi}_{h}^{(j)}$, i.e.,
\begin{equation*}
\boldsymbol{\psi}_{h}^{(j)}\sim
Dirichlet(\beta_{j0},\cdots,\beta_{jd_j}),
\end{equation*}
for $h=1,\cdots,k$ and
$j=1,\cdots,p$. If the sample size $n$ is small, we suggest using a
flat and weak prior by setting $\beta_{j0}=\cdots= \beta_{jd_{j}}=1$ for
all $j$. For the Dirichlet prior in Equation~\ref{eq3}, \cite{teh2006hierarchical} suggested that
the hyper-parameter $\alpha$ should be a small number, thus we set $\alpha=0.25$ by default. In a specific application where users have more knowledge about the concentration level, the value of $\alpha$ can be changed according to the prior knowledge.

Since $x_{ij}$'s are independent conditional on the latent component
$z_{i}$'s and $\boldsymbol{\psi}_{z_i}^{(j)}$'s as in
Equation~\ref{eq4}, the joint likelihood of $\boldsymbol{x}$ given
$\boldsymbol{z}$ and $\Psi$ can be written as
\begin{equation*}
p(\boldsymbol{x}|\boldsymbol{z},\Psi)=\prod_{i=1}^n\prod_{j=1}^p
p(x_{ij}|z_i,\boldsymbol{\psi}_{z_i}^{(j)})=\prod_{i=1}^n\prod_{j=1}^p
\psi_{z_ix_{ij}}^{(j)}.
\end{equation*}
Since $\boldsymbol{z}$ and $\Psi$ are
independent, the augmented joint posterior distribution is
$p(\boldsymbol{z},\Psi|\boldsymbol{x})\propto
p(\boldsymbol{x}|\boldsymbol{z},\Psi)p(\boldsymbol{z}) p(\Psi)$.

To sample from the joint posterior distribution, we use the
following Gibbs sampler algorithm \ref{alg1}:

\begin{algorithm}[!htbp]
\caption{Gibbs Sampler Algorithm}\label{alg1}
\begin{algorithmic}[1]
\State{Initialize parameters $\boldsymbol{z}$ and $\boldsymbol{\Psi}$.
 For convenience, we can set all samples in different components initially. Set $z_i=i$, $i=1,\cdots,n$ and $k=n$. Set $\boldsymbol{\psi}_{i}^{(j)}$ by sampling
 from $\boldsymbol{\psi}_{i}^{(j)}\sim Dirichlet(\beta_{j0},\cdots,\beta_{jd_j})$, for $i=1,\cdots,n$ and $j=1,\cdots,p$.}
\State{Update $z_i$, $i=1,\cdots,n$, by sampling according to the following weights:
     \begin{equation*}
\begin{split}
&P(z_i=h|\boldsymbol{z}_{-i},\boldsymbol{x},\Psi)\propto\frac{n_{h,-i}}{n+\alpha-1}\prod_{j=1}^p\psi_{hx_{ij}}^{(j)},~~h=1,\cdots,k;\\
&P(z_i=k+1|\boldsymbol{z}_{-i},\boldsymbol{x},\Psi)\\
&~~~~~~~~~~~\propto\frac{\alpha}{n+\alpha-1}\prod_{j=1}^p\frac{\prod_{c=0}^{d_j}\Gamma(I(x_{ij}=c)+\beta_{jc})}{\Gamma(\sum_{c=0}^{d_j}\beta_{jc}+1)}.\\
\end{split}
\end{equation*}
If $z_i=k+1$, we obtain a new component and increase $k$ by 1.
Then we sample a new component parameter $\boldsymbol{\psi}_{k+1}$
from
$\boldsymbol{\psi}_{k+1}^{(j)}\sim
Dirichlet(I(x_{ij}=0)+\beta_{j0},\cdots,I(x_{ij}=d_j)+\beta_{jd_j})$,
for $j=1,\cdots,p$.}
\State{Sort the components decreasingly according to the number of samples in each component. Delete empty components and re-calculate $k$.}
\State{Update $\boldsymbol{\psi}_{h}^{(j)}$ by sampling from
\begin{equation*}
\begin{split}
\boldsymbol{\psi}_{h}^{(j)}&\sim Dirichlet(\sum_{i:z_i=h}I(x_{ij}=0)+\beta_{j0},\\
&~~~~~~~~~~~~~~~~~~~~~~~~~\cdots,\sum_{i:z_i=h}I(x_{ij}=d_j)+\beta_{jd_j}),
\end{split}
\end{equation*}
for $h=1,\cdots,k$ and $j=1,\cdots,p$.}
\State{Repeat Step 2-4 until convergence.}
\State{Re-scale the multinomial probabilities without the category for missingness:
\begin{equation*}
\tilde{\psi}_{hc}^{(j)}={\psi_{hc}^{(j)}}/{(1-\psi_{h0}^{(j)})},
\end{equation*}
for $h=1,\cdots,k$; $j=1,\cdots,p$; and $c=1,\cdots,d_j$.}
\end{algorithmic}
\end{algorithm}

In Step 2, for $h=1,\cdots,k$, the conditional posterior distribution is derived as follows:
\begin{equation*}
\begin{split}
p(z_i&=h|\boldsymbol{z}_{-i},\boldsymbol{x},\Psi)=
p(z_i=h|\boldsymbol{z}_{-i},\boldsymbol{x}_i,\boldsymbol{\psi}_h)\\
&\propto p(z_i=h|\boldsymbol{z}_{-i})p(\boldsymbol{x}_i|z_i=h,\boldsymbol{\psi}_h)\\
&=\frac{n_{h,-i}}{n+\alpha-1}\prod_{j=1}^pp(x_{ij}|z_i=h,\boldsymbol{\psi}_h^{(j)})\\
&=\frac{n_{h,-i}}{n+\alpha-1}\prod_{j=1}^p\psi_{hx_{ij}}^{(j)}.\\
\end{split}
\end{equation*}
For $h=k+1$, we have:
\begin{equation*}
\begin{split}
p(z_i&=k+1|\boldsymbol{z}_{-i},\boldsymbol{x},\Psi)\propto p(z_i=k+1|\boldsymbol{z}_{-i})p(\boldsymbol{x}_i|z_i=k+1)\\
&=\frac{\alpha}{n+\alpha-1}\prod_{j=1}^pp(x_{ij}|z_i=k+1),\\
\end{split}
\end{equation*}
where
\begin{equation*}
\begin{split}
p(x_{ij}|z_i&=k+1)=\int p(x_{ij}|z_i=k+1,\boldsymbol{\psi}_{k+1}^{(j)})p(\boldsymbol{\psi}_{k+1}^{(j)})d\boldsymbol{\psi}_{k+1}^{(j)}\\
&=\int\prod_{c=0}^{d_j}(\psi_{k+1,c}^{(j)})^{I(x_{ij}=c)}\cdot\prod_{c=0}^{d_j}(\psi_{k+1,c}^{(j)})^{\beta_{jc}-1}d\boldsymbol{\psi}_{k+1}^{(j)}\\
&=\frac{\prod_{c=0}^{d_j}\Gamma(I(x_{ij}=c)+\beta_{jc})}{\Gamma(\sum_{c=0}^{d_j}\beta_{jc}+1)}.\\
\end{split}
\end{equation*}

In Step 4, the conditional posterior distribution comes from:
\begin{equation*}
\begin{split}
p(\boldsymbol{\psi}_{h}^{(j)}|&\boldsymbol{x},\boldsymbol{z}) \propto p(\boldsymbol{x}|\boldsymbol{z},\boldsymbol{\psi}_{h}^{(j)})
p(\boldsymbol{\psi}_{h}^{(j)})\\
&\propto \prod_{i:z_i=h}\prod_{c=0}^{d_j}
(\psi_{hc}^{(j)})^{I(x_{ij}=c)}\cdot \prod_{c=0}^{d_j}(\psi^{(j)}_{hc})^{\beta_{jc}-1}\\
&=\prod_{c=0}^{d_j}
\big[\psi_{hc}^{(j)}\big]^{\sum_{i:z_i=h}I(x_{ij}=c)+\beta_{jc}-1}.\\
\end{split}
\end{equation*}

The above Gibbs sampler algorithm is implemented in the R package \texttt{MMDai}.

\subsection{Identifiability}\label{sec:indentifiable}
\cite{dunson2009nonparametric} proved that any multivariate
categorical distribution can be decomposed to a finite mixture
product-multinomial model for some $k$. It should be noted that
this decomposition is not identifiable if no restrictions are
placed, where ``identifiable'' means the decomposition is unique
up to label-switching. In fact, \cite{elmore2003identifiability}
proved that the $k$-component finite mixture of univariate
multinomial distribution is identifiable if and only if the number
of trials $m$ in multinomial distributions satisfies the condition
that $m>2k-1$. For the multivariate multinomial distribution, we
can also prove that it is identifiable if and only if $m_j>2k-1$
for all $j$. In most real applications, we only have $m_j=1$ and
thus the decomposition is always unidentifiable.

Fortunately, identifiability is not a problem in our analysis.
Firstly, we are only interested in the estimation of the joint
distribution, which is unique though the decomposition is not. The
main issue here is whether the joint distribution can be
approximated well enough. \cite{vermunt2008multiple} discussed the
above difference between estimating joint distribution and
clustering when the latent class model is adopted. Secondly, with
the Dirichlet prior in our Bayesian approach, different
decompositions are weighted unequally and the simpler model is
preferred. For the multiple non-identifiable decompositions, our
algorithm usually converges to the decomposition with the smallest
$k$.

\section{Simulations}
\label{sec:sim}

In this section, we check the performance of our method and compare it with other methods based on synthetic data. In each simulation experiment, we first synthesize a set of complete
data following a given data model, then mask a certain percentage of the observations according to the corresponding missing mechanism to generate the incomplete dataset as our input data, finally use the masked entries to test the performance of statistical inference and imputation. We consider the following missing mechanisms in simulation studies:
\begin{itemize}
 \item MCAR: The missingness does not depend on observed and missing values,
     \begin{equation*}
     p(x_{ij}=0)=0.2,
     \end{equation*}
     for all $i=1,\cdots,n$ and $j=1,\cdots,p$.

 \item MAR: The first variable is fully observed and the missingness in other variables depends on the observation in the first variable:
     \begin{equation*}
     p(x_{ij}=0|x_{i1}=1)=0.1,~~p(x_{ij}=0|x_{i1}=2)=0.3,
     \end{equation*}
     for $i=1,\cdots,n$ and $j=2,\cdots,p$.

 \item MNAR: The missingness depends on the missing value itself:
     \begin{equation*}
     p(x_{ij}=0|x_{ij}=1)=0.1,~~p(x_{ij}=0|x_{ij}=2)=0.3,
     \end{equation*}
     for all $i=1,\cdots,n$ and $j=1,\cdots,p$.
\end{itemize}

Considering the uncertainty of Monte Carlo simulations, we repeat the
experiments, including the data synthesis step, for 100 times independently for each data model. To compare the performance with existing methods, we also apply the DPMPM model in \cite{si2013nonparametric} and MICE on these datasets. DPMPM and MICE represent two typical ideas for imputing incomplete categorical data respectively: joint modeling by Bayesian latent class model and FCS by iterative regression. See the supplementary materials for the detailed R code.

\subsection{Case 1: Data from Mixture Model}\label{sec:mix}
In this case, we generate binary data ($d_j=2$ for all $j$) from a finite mixture of product-multinomial distributions with $n=50$ and $p=20$. The true parameters in the mixture model are set as follows: the number of components $k=3$, $\boldsymbol{\Theta}$ and $\boldsymbol{\Psi}$ are sampled from $\boldsymbol{\Theta}\sim Dirichlet(10,\cdots,10)$ and $\boldsymbol{\psi}^{(j)}_h\sim Dirichlet(0.5,\cdots,0.5)$ for $h=1,\cdots,k$ and $j=1,\cdots,p$, respectively.

\begin{figure}[!htbp]
  \centering
  \includegraphics[width=9cm]{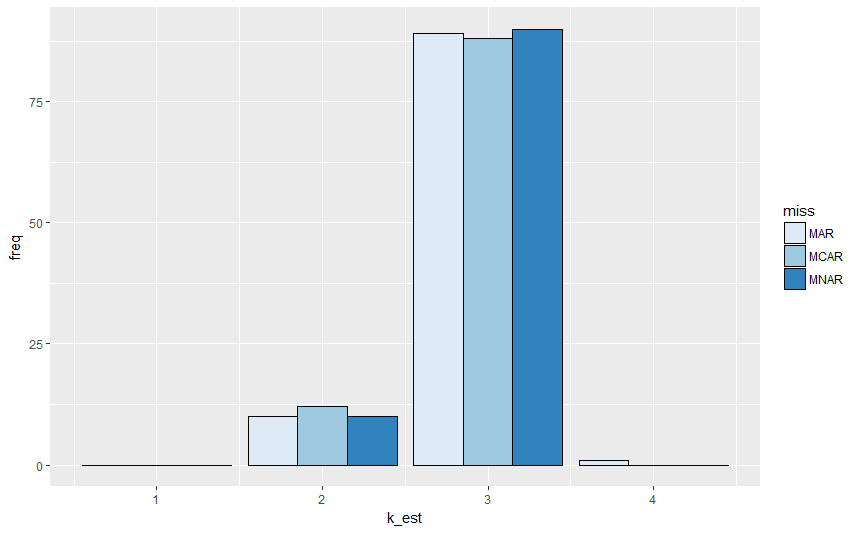}\\
  \caption{Histograms of the estimated $k$ in 100 replications under different missing mechanisms.}\label{Fig1}
\end{figure}

First we show that DPMCPM can identify the correct number of components in the mixture model. Figure~\ref{Fig1} presents the histogram of the estimated $k$ in 100 replications under different missing mechanisms. In most cases, DPMCPM can capture the true number of components $k=3$. In other cases, DPMCPM may select the simpler model to interpret datasets.

Then we show that DPMCPM can outperform DPMPM and MICE in the imputation of missing values, where the imputation accuracy is defined as follows:
\begin{equation*}
\text{imputation accuracy}=\frac{\text{the number of correct imputation}}{\text{the number of masked entries}}.
\end{equation*}
Here the number of correct imputation is calculated based on the true complete data. Table~\ref{table2} presents the mean and standard error of imputation accuracy of the three methods for 100 replications under different missing mechanisms. It shows that DPMCPM significantly outperformed other methods.

\begin{table}[!htbp]
\centering
\begin{tabular}{cccc}
  \toprule
    & DPMCPM & DPMPM & MICE \\
  \midrule
 MCAR & 0.7860(0.044) & 0.7031(0.040) & 0.5746(0.039)\\
 MAR  & 0.7744(0.049) & 0.6987(0.048) & 0.5687(0.052) \\
 MNAR & 0.7684(0.050) & 0.6811(0.049) & 0.5752(0.044)\\
 \bottomrule
\end{tabular}
\caption{Mean and standard error of imputation accuracy of the three methods under different missing mechanisms. Data are generated from the finite mixture product-multinomial model.}\label{table2}
\end{table}

Besides the imputation accuracy, we can also compare the performance on other statistical inference problems. From DPMCPM, we obtain an estimation of the multivariate multinomial distribution with the estimates of $\boldsymbol{\Theta}$ and $\boldsymbol{\Psi}$. Although the distribution estimation can be a little different from the true distribution due to the existence of missingness and the limited sample size, DPMCPM can perform more or better statistical inference than DPMPM and MICE based on the estimated distribution. Table~\ref{table3} presents the gap between the estimated correlation matrix and the true correlation matrix. The gap is defined as follows:
\begin{equation*}
\text{gap}=\sum_i\sum_j(\hat\sigma_{ij}-\sigma_{ij})^2,
\end{equation*}
where $\hat\sigma_{ij}$ and $\sigma_{ij}$ are the estimated and true correlation between the $i$-th and $j$-th variables, respectively. The mean and standard error of the gap for 100
replications are summarized in Table~\ref{table3}. Because MICE can not provide an estimation of the distribution directly, we use instead the sample distribution from its imputation. The table shows that DPMCPM have better correlation estimation than other methods under all missing mechanisms.
\begin{table}[!htbp]
\centering
\begin{tabular}{cccc}
  \toprule
    & DPMCPM & DPMPM  & MICE\\
  \midrule
 MCAR & 7.5968(2.216) & 9.2249(2.656) & 11.8448(2.088)\\
 MAR  & 7.8092(2.366) & 8.8777(2.279) & 12.1515(2.487)\\
 MNAR & 7.1693(2.287) & 9.3119(2.549) &
 11.7781(2.395)\\
 \bottomrule
\end{tabular}
\caption{Mean and standard error of the gap between the estimated and true correlation matrix under different missing mechanisms.}\label{table3}
\end{table}

\subsection{Case 2: Nonlinear Association}\label{sec:nonlinear}
In addition to the data generated from mixture models, we also perform simulation studies on more general cases. In practice, it is cumbersome to design and simulate
general categorical data distribution with large $p$ due to the exponentially increasing size of the contingency table. Here we design a simple nonlinear experiment to demonstrate the power of DPMCPM in this aspect.

As an example, we generate three Bernoulli random variables, $V_1$, $V_2$ and $V_3$, as follows. Assume that $V_1\sim Bernoulli(0.3)$ and $V_2\sim Bernoulli(0.5)$ are independent, where $V\sim Bernoulli(p)$ denotes that $V$ follows a Bernoulli distribution with probability $Pr(V=1)=p$. Let $V_3$ be the output of the exclusive-or operator on $V_1$ and $V_2$ with probability 95\% and be an independent Bernoulli random error, i.e.,
$V_3\sim Bernoulli(0.5)$, with probability 5\%. Essentially, this is a data model with strong nonlinear association. We set $n=300$, $p=3$ and the same missing mechanisms as in Section~\ref{sec:mix}. The mean
and standard error of the imputation accuracy in this study are summarized in Table~\ref{table4}. The results
show that MICE does not capture the nonlinear association among variables while DPMCPM does it well.  DPMCPM also performs better than DPMPM here.
\begin{table}[!htbp]
\centering
\begin{tabular}{cccc}
  \toprule
    & DPMCPM & DPMPM & MICE \\
  \midrule
 MCAR  & 0.8527(0.031) & 0.7527(0.052) & 0.5644(0.038) \\
 MAR  & 0.8699(0.041) & 0.7832(0.058) & 0.5054(0.055) \\
 MNAR & 0.7935(0.060) & 0.6945(0.072) &
 0.5179(0.041)\\
 \bottomrule
\end{tabular}
\caption{Mean and standard error of the imputation accuracy of three methods under different missing mechanisms. Data are generated from the nonlinear association model.}\label{table4}
\end{table}

\section{Application on Recommendation System Problem}\label{sec:app}
In this section, we present a real application on a recommendation system problem. \cite{harper2016movielens}
contributed a dataset about the ratings of movies. The entire
ratings table in \cite{harper2016movielens} contains 20000263
ratings on 27278 movies from 138493 users. In real applications of missing data, the true values of missingness are always unknown.  For the purpose of
cross-validation, we extract a subset of data with low missingness
such that we mask a high percent of them for evaluating the
imputation accuracy. First, we select the movies which were rated
by more than 25\% users. Then we remain the users who rated more
than 95\% of the selected movies. The resulting dataset contains
68861 ratings on 38 movies from 1837 users, where the percentage
of missingness is 1.35\%. The data matrix \emph{MovieRate} in the
\texttt{MMDai} package is the resulted dataset.

Since the original ratings in \cite{harper2016movielens} are
ordinal data made on a 5-star scale with half-star increments (0.5
stars - 5.0 stars), we consider two plans: (1) transform to binary data according to a cutoff (thus the data is pure categorial); (2) round up to 5-category data (thus the data is ordinal). To evaluate the performance of DPMCPM, we mask
40\% of the ratings in \emph{MovieRate} under MCAR, resulting in the incomplete
``observed data'', where the percentage of
missing values is 40.81\%. Predicting the unobserved favor of a
user on a particular movie is equivalent to impute a missing value
in incomplete data.

The analyses in Section~\ref{sec:sim} are repeated on this real dataset. As an example of recommendation system, we also compare with the classical SVD approximation algorithm for recommendation problems. The data \emph{MovieRate} is masked and imputed for 100 times independently. The mean and standard error of the imputation accuracy are summarized in Table~\ref{table5}. To check the
effect of the cutoff used to dichotomize the original ordinal
data, we repeat the experiments by varying the cutoff from 4.0
to 3.5 and 3.0. Table~\ref{table5} shows that DPMCPM has
significantly higher imputation accuracy than other approaches on
\emph{MovieRate}. The performance improvement is not sensitive to
the choice of the cutoff. The improved accuracy can be very
helpful for reducing the cost of bad recommendations in real
life.

\begin{table*}[!htbp]
\centering
\begin{tabular}{ccccc}
  \toprule
   Cutoff & DPMCPM & DPMPM & MICE & SVD\\
   \midrule
   Binary (Cutoff-4.0) & 0.7598(0.002)&¡¡0.6805(0.003) &  0.6859(0.002) &0.6765(0.004)\\
   Binary (Cutoff-3.5) & 0.8158(0.002) & 0.7424(0.002)&  0.7400(0.002) &0.7413(0.003)\\
   Binary (Cutoff-3.0) & 0.9075(0.001) & 0.8602(0.002) &  0.8445(0.002) &  0.8599(0.002)\\
   5-Categoty & 0.5480(0.003)	& 0.4321(0.003) &	0.4198(0.002) & 0.4449(0.005)\\
  \bottomrule
\end{tabular}
\caption{Imputation accuracy comparison on the movie data.}\label{table5}
\end{table*}

\begin{figure}[!htbp]
  \centering
  \includegraphics[width=9cm]{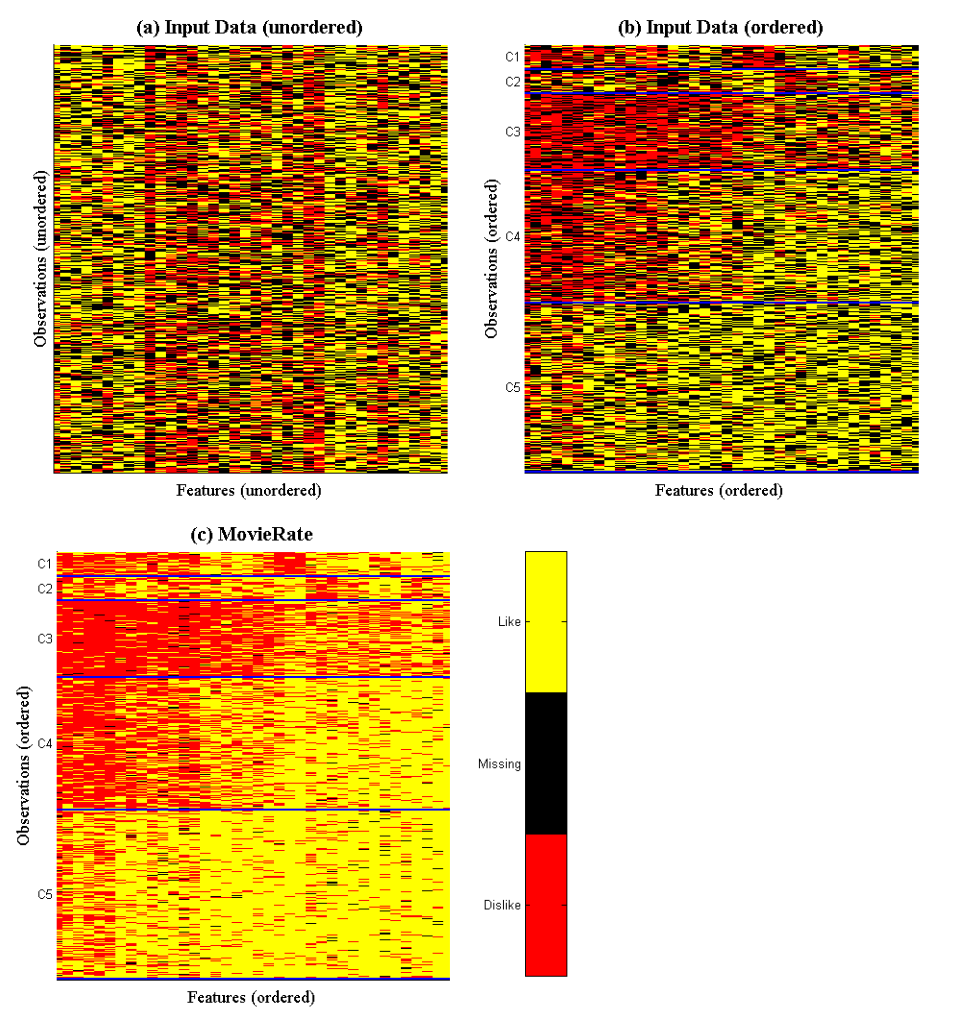}
  \caption{Heatmap plots of an unordered input Data, the input data ordered according to the model fitting by DPMCPM, and the true data \emph{MovieRate} with the same ordering.}\label{fig1}
\end{figure}

Except for the imputation of missing values, we also explore the
clustering structure among movies according to the fitting of DPMCPM. Figure~\ref{fig1} shows the heatmap plots of an input data from one of the
100 experiments (both unordered and ordered versions) and the true
data \emph{MovieRate}, where the rows correspond to users and the
columns correspond to movies. The order is based on the latent class inferred by DPMCPM. The figure demonstrated that
DPMCPM could detect the cluster structure efficiently despite
the high missingness in the input data.

To demonstrate the usage of the estimated distribution from DPMCPM, we perform the Fisher exact test to test the independence of each pair of movies according to the estimated joint distribution. The five pairs with the least
and largest $p$-values are reported in Table~\ref{table6}. The
association results agree with common sense based on the nature of
these movies.
\begin{table}[!htbp]
\tiny
\centering
\begin{tabular}{c|c|c}
  \hline
  Movie 1 & Movie 2 & $p$-value \\
  \hline
  Star Wars: Episode IV (1977)
 & Star Wars: Episode V (1980)
 & 0.00004 \\
  Star Wars: Episode V (1980)
 & Star Wars: Episode VI (1983)
 & 0.00023 \\
 Star Wars: Episode IV (1977)
  & Star Wars: Episode VI (1983)
 & 0.00069\\
 The Fugitive (1993)
 & Jurassic Park (1993)
 & 0.00086 \\
 Jurassic Park (1993)
 & Men in Black (1997)
 & 0.00106 \\
  Twelve Monkeys (1995)
 & Ace Ventura: Pet Detective (1994)
 & 1.00000 \\
 Apollo 13 (1995)
 & Fight Club (1999)
 & 1.00000 \\
 Pulp Fiction (1994)
  & Mission: Impossible (1996)
 & 1.00000 \\
  Pulp Fiction (1994)
 & Independence Day (1996)
 & 1.00000 \\
  Speed (1994)
 & Fight Club (1999)
 & 1.00000 \\
\hline
\end{tabular}
\caption{The pairs of movies with the least and largest
Fisher test $p$-values.}\label{table6}
\end{table}

\section{Conclusion}\label{sec:summary}

In this paper, we introduce the DPMCPM method,
which implemented multivariate multinomial distribution
estimation for categorical data with missing
values. DPMCPM inherits some nice properties of DPMPM. It
avoids the arbitrary setting of the number of mixture components
by using the Dirichlet process. Also, DPMCPM is applicable
for any categorical data regardless of the true distribution
and can capture the complex association among variables.
Unlike DPMPM, DPMCPM can model general missing mechanisms and handle high missingness efficiently, thus DPMCPM can achieve more accurate results in empirical studies. Through simulation studies and a real application, we demonstrate
that DPMCPM performs better statistical inference and
imputation than other methods.

It shall be noted that approximating a general distribution
of high-dimensional categorical data is still a hard task,
although DPMCPM could approximate the true general
distribution at arbitrary precision theoretically if $n$ is big
enough. To estimate the underlying general distribution
accurately, the required sample size $n$ still increases
exponentially with the number of variables $p$. Thus our method may be limited when $p$ grows with $n$. Fortunately, in
many applications, there is underlying cluster structures in the
dataset. In these cases, the underlying  distribution is
close to a mixture of product-multinomials, thus the sample
size needed for DPMCPM to estimate the underlying
distribution accurately is much smaller. For example, in
recommendation system problems, it is reasonable to assume that the
customers with similar historical activities have similar tastes.
For handling such problems, DPMCPM would have more advantages
than traditional approaches. Future work shall try to improve the efficiency for the case that there is no underlying cluster structures in the high-dimensional dataset.

\appendix
\section{Proof}
\subsection{Theorem 1}
\label{appA}

\emph{Proof:} Here is a constructive proof for Theorem 1.

Assume that there are $p$ variables and the $j$-th variable has $d_j$ categories. Let $\pi_{c_1,\cdots,c_p}=p(X_{1}=c_1,\cdots,X_{p}=c_p)$ then $\sum_{c_1=1}^{d_1}\cdots\sum_{c_p=p}^{d_p}\pi_{c_1,\cdots,c_p}=1$.

For all missing mechanisms belong to MCAR, MAR or MNAR family, the missing rate can be written as $p(\boldsymbol{r}|\boldsymbol{x})=\prod_{i=1}^n\prod_{j=1}^pp(r_{ij}|\boldsymbol{x}_i)$, which is a function of $\boldsymbol{x}_i$. For each variable, let
\begin{equation*}
\begin{split}
q_{c_1,\cdots,c_p}^{(j)}&=p(r_{ij}=0|x_{i1}=c_1,\cdots,x_{ip}=c_p),\\
\end{split}
\end{equation*}

For any joint distribution $\boldsymbol{\Pi}$ with missing mechanism $\textbf{Q}$ where $\textbf{Q}=\{q_{c_1,\cdots,c_p}^{(j)}\}$ is a table with $d_1\times\cdots\times d_p\times p$ cells, we can estimate a new joint probability table by adding an extra category to denote missingness,
\begin{equation*}
\boldsymbol{\tilde\Pi}=\tilde\theta_1\cdot\boldsymbol{\tilde\psi}_1^{(1)}
\otimes\cdots\otimes\boldsymbol{\tilde\psi}_1^{(p)}+\cdots+\tilde\theta_{k}\cdot\boldsymbol{\tilde\psi}_{k}^{(1)}
\otimes\cdots\otimes\boldsymbol{\tilde\psi}_{k}^{(p)}
\end{equation*}
where $\boldsymbol{\tilde\psi}_h^{(j)}=(\tilde\psi_{h0}^{(j)},\tilde\psi_{h1}^{(j)},\cdots,\tilde\psi_{hd_j}^{(j)})$.

The problem we need to prove is, there exists a set of parameters $\{\boldsymbol{\tilde\Theta},\boldsymbol{\tilde\Psi}\}$ that can recover the original joint distribution table $\boldsymbol{\Pi}$ and missing mechanism $\textbf{Q}$ after scaling the extra category in the latent classes.
In another word, given $\boldsymbol{\Pi}$ and $\textbf{Q}$, there exists a set of solution $\{\boldsymbol{\tilde\Theta},\boldsymbol{\tilde\Psi}\}$ under our estimates.

Note that for any general cases, we always have following decomposition
\begin{equation*}
\begin{split}
\boldsymbol{\Pi}=\sum_{c_1,\cdots,c_p}\pi_{c_1,\cdots,c_p}\cdot I_{c_1}\otimes\cdots\otimes I_{c_p}.\\
\end{split}
\end{equation*}

Define a series of latent classes $\mathcal{H}=\{\overline{c_1,\cdots,c_p}\}$ where $\overline{c_1,\cdots,c_p}$ is a notation of latent class. Here we design a set of parameters $\{\boldsymbol{\tilde\Theta},\boldsymbol{\tilde\Psi}\}$:
\begin{equation*}
\begin{split}
\boldsymbol{\tilde\Theta}&=\{\tilde\theta_h:h\in\mathcal{H}\},~~\text{where}~~\tilde\theta_{\overline{c_1,\cdots,c_p}}=\pi_{c_1,\cdots,c_p},\\
\boldsymbol{\tilde\Psi}&=\{\boldsymbol{\tilde\psi}_{h}^{(j)}:h\in\mathcal{H}\},
\end{split}
\end{equation*}
where $\boldsymbol{\tilde\psi}_{h}^{(j)}=(\tilde\psi_{h0}^{(j)},\tilde\psi_{h1}^{(j)}\cdots,\tilde\psi_{hd_j}^{(j)})$ is a $d_j+1$-dimension vector. For the vector $\boldsymbol{\tilde\psi}_{\overline{c_1,\cdots,c_p}}^{(j)}$, we have $\tilde\psi_{\overline{c_1,\cdots,c_p}0}^{(j)}=q_{c_1,\cdots,c_p}^{(j)}$ and $\tilde\psi_{\overline{c_1,\cdots,c_p}c_j}^{(j)}=1-q_{c_1,\cdots,c_p}^{(j)}$ with other elements are 0.

Then we re-scale the extra category $\boldsymbol{\psi}_{h}^{(j)}=(\psi_{h1}^{(j)},\cdots,\psi_{hd_j}^{(j)})$  where $\psi_{hc}^{(j)}=\tilde\psi_{hc}^{(j)}/(1-\tilde\psi_{h0}^{(j)})$.
Thus, we have $\boldsymbol{\psi}_{\overline{c_1,\cdots,c_p}}^{(j)}=I_{c_j}$ where $I_{c_j}$ is a vector with length $d_j$ that the $c_j$-th element is 1 and other elements are 0.

Under this $\{\boldsymbol{\tilde\Theta},\boldsymbol{\tilde\Psi}\}$ setting, the joint distribution is
\begin{equation*}
\begin{split}
&\sum_{h\in \mathcal{H}}\tilde\theta_{h}\cdot\boldsymbol{\psi}_{h}^{(1)}
\otimes\cdots\otimes\boldsymbol{\psi}_{h}^{(p)}\\
=&\sum_{\overline{c_1,\cdots,c_p}}\pi_{c_1,\cdots,c_p}\cdot I_{c_1}\otimes\cdots\otimes I_{c_p}=\boldsymbol{\Pi}.\\
\end{split}
\end{equation*}

In this decomposition, the posterior probability of latent class
\begin{equation*}
\begin{split}
&p(z_i=\overline{c_1,\cdots,c_p}|x_{i1}=c_1,\cdots,x_{ip}=c_p)\\
=~&p(z_i=\overline{c_1,\cdots,c_p},x_{i1}=c_1,\cdots,x_{ip}=c_p)\\
&~~~~~~/p(x_{i1}=c_1,\cdots,x_{ip}=c_p)\\
=~&\pi_{c_1,\cdots,c_p}/\pi_{c_1,\cdots,c_p}=1.
\end{split}
\end{equation*}

For the missing mechanism, on the level of latent class, we have
\begin{equation*}
\begin{split}
&p(r_{ij}=1|z_i=\overline{c_1,\cdots,c_p},x_{i1}=c_1,\cdots,x_{ip}=c_p)=\tilde\psi_{\overline{c_1,\cdots,c_p}c_j}^{(j)},\\
\end{split}
\end{equation*}
then
\begin{equation*}
\begin{split}
&p(r_{ij}=0|z_i=\overline{c_1,\cdots,c_p},x_{i1}=c_1,\cdots,x_{ip}=c_p)\\
=~&1-p(r_{ij}=1|z_i=\overline{c_1,\cdots,c_p},x_{i1}=c_1,\cdots,x_{ip}=c_p)\\
=~&1-\tilde\psi_{\overline{c_1,\cdots,c_p}c_j}^{(j)}=\tilde\psi_{\overline{c_1,\cdots,c_p}0}^{(j)}
\end{split}
\end{equation*}

Thus, integrating the level of latent class, we have
\begin{equation*}
\begin{split}
&p(r_{ij}=0|x_{i1}=c_1,\cdots,x_{ip}=c_p)\\
=~&\sum_{h\in \mathcal{H}}p(r_{ij}=0|z_i=h,x_{i1}=c_1,\cdots,x_{ip}=c_p)\\
&~~\cdot p(z_i=h|x_{i1}=c_1,\cdots,x_{ip}=c_p)\\
=~&p(r_{ij}=0|z_i=\overline{c_1,\cdots,c_p},x_{i1}=c_1,\cdots,x_{ip}=c_p)\\
=~&\tilde\psi_{\overline{c_1,\cdots,c_p}0}^{(j)}=q_{c_1,\cdots,c_p}^{(j)}.\\
\end{split}
\end{equation*}

So we show that any joint distribution with general missing mechanism can be captured by our latent class model. Proven.

\textbf{Remark:} This proof is a constructive proof. We show that there exists a construction of the latent class model that can capture any general missing mechanisms under any joint distributions. In specific cases, this construction may be not unique since decomposition of joint distribution is multiple, and not optimal in terms of complexity of model. Actually, this proof shows that the estimate can always attain the true joint distribution with right missing mechanism, which is the maximum a posteriori estimate when sample size $n$ is large enough.

\bibliographystyle{imsart-nameyear}
\bibliography{reference}

\begin{thebibliography}{32}

\bibitem[\protect\citeauthoryear{Audigier, Husson and
  Josse}{2017}]{audigier2017mimca}
\begin{barticle}[author]
\bauthor{\bsnm{Audigier},~\bfnm{Vincent}\binits{V.}},
  \bauthor{\bsnm{Husson},~\bfnm{Fran{\c{c}}ois}\binits{F.}} \AND
  \bauthor{\bsnm{Josse},~\bfnm{Julie}\binits{J.}}
(\byear{2017}).
\btitle{MIMCA: Multiple imputation for categorical variables with multiple
  correspondence analysis}.
\bjournal{Statistics and Computing}
\bvolume{27}
\bpages{501--518}.
\end{barticle}
\endbibitem

\bibitem[\protect\citeauthoryear{Donders et~al.}{2006}]{donders2006review}
\begin{barticle}[author]
\bauthor{\bsnm{Donders},~\bfnm{A~Rogier~T}\binits{A.~R.~T.}},
  \bauthor{\bparticle{van~der} \bsnm{Heijden},~\bfnm{Geert~JMG}\binits{G.~J.}},
  \bauthor{\bsnm{Stijnen},~\bfnm{Theo}\binits{T.}} \AND
  \bauthor{\bsnm{Moons},~\bfnm{Karel~GM}\binits{K.~G.}}
(\byear{2006}).
\btitle{Review: A gentle introduction to imputation of missing values}.
\bjournal{Journal of Clinical Epidemiology}
\bvolume{59}
\bpages{1087--1091}.
\end{barticle}
\endbibitem

\bibitem[\protect\citeauthoryear{Dunson and
  Xing}{2012}]{dunson2009nonparametric}
\begin{barticle}[author]
\bauthor{\bsnm{Dunson},~\bfnm{David~B}\binits{D.~B.}} \AND
  \bauthor{\bsnm{Xing},~\bfnm{Chuanhua}\binits{C.}}
(\byear{2012}).
\btitle{Nonparametric Bayes modeling of multivariate categorical data}.
\bjournal{Journal of the American Statistical Association}
\bvolume{104}
\bpages{1042--1051}.
\bdoi{10.1198/jasa.2009.tm08439}
\end{barticle}
\endbibitem

\bibitem[\protect\citeauthoryear{Elmore and
  Wang}{2003}]{elmore2003identifiability}
\begin{btechreport}[author]
\bauthor{\bsnm{Elmore},~\bfnm{Ryan~T}\binits{R.~T.}} \AND
  \bauthor{\bsnm{Wang},~\bfnm{Shaoli}\binits{S.}}
(\byear{2003}).
\btitle{Identifiability and estimation in finite mixture models with
  multinomial coefficients}
\btype{Technical Report},
\bpublisher{03-04, Penn State University}.
\end{btechreport}
\endbibitem

\bibitem[\protect\citeauthoryear{Formann}{2007}]{formann2007mixture}
\begin{barticle}[author]
\bauthor{\bsnm{Formann},~\bfnm{Anton~K}\binits{A.~K.}}
(\byear{2007}).
\btitle{Mixture analysis of multivariate categorical data with covariates and
  missing entries}.
\bjournal{Computational Statistics \& Data Analysis}
\bvolume{51}
\bpages{5236--5246}.
\bdoi{10.1016/j.csda.2006.08.020}
\end{barticle}
\endbibitem

\bibitem[\protect\citeauthoryear{Harper and
  Konstan}{2016}]{harper2016movielens}
\begin{barticle}[author]
\bauthor{\bsnm{Harper},~\bfnm{F~Maxwell}\binits{F.~M.}} \AND
  \bauthor{\bsnm{Konstan},~\bfnm{Joseph~A}\binits{J.~A.}}
(\byear{2016}).
\btitle{The movielens datasets: History and context}.
\bjournal{ACM Transactions on Interactive Intelligent Systems (TiiS)}
\bvolume{5}
\bpages{19}.
\end{barticle}
\endbibitem

\bibitem[\protect\citeauthoryear{Hicks et~al.}{2018}]{hicks2017missing}
\begin{barticle}[author]
\bauthor{\bsnm{Hicks},~\bfnm{Stephanie~C}\binits{S.~C.}},
  \bauthor{\bsnm{Townes},~\bfnm{F~William}\binits{F.~W.}},
  \bauthor{\bsnm{Teng},~\bfnm{Mingxiang}\binits{M.}} \AND
  \bauthor{\bsnm{Irizarry},~\bfnm{Rafael~A}\binits{R.~A.}}
(\byear{2018}).
\btitle{Missing data and technical variability in single-cell RNA-sequencing
  experiments}.
\bjournal{Biostatistics}
\bvolume{19}
\bpages{562-578}.
\end{barticle}
\endbibitem

\bibitem[\protect\citeauthoryear{Hu et~al.}{2018}]{hu2018dirichlet}
\begin{barticle}[author]
\bauthor{\bsnm{Hu},~\bfnm{Jingchen}\binits{J.}},
  \bauthor{\bsnm{Reiter},~\bfnm{Jerome~P}\binits{J.~P.}},
  \bauthor{\bsnm{Wang},~\bfnm{Quanli}\binits{Q.}} \betal{et~al.}
(\byear{2018}).
\btitle{Dirichlet process mixture models for modeling and generating synthetic
  versions of nested categorical data}.
\bjournal{Bayesian Analysis}
\bvolume{13}
\bpages{183--200}.
\end{barticle}
\endbibitem

\bibitem[\protect\citeauthoryear{Jolani et~al.}{2015}]{jolani2015imputation}
\begin{barticle}[author]
\bauthor{\bsnm{Jolani},~\bfnm{Shahab}\binits{S.}},
  \bauthor{\bsnm{Debray},~\bfnm{Thomas~PA}\binits{T.~P.}},
  \bauthor{\bsnm{Koffijberg},~\bfnm{Hendrik}\binits{H.}},
  \bauthor{\bparticle{van} \bsnm{Buuren},~\bfnm{Stef}\binits{S.}} \AND
  \bauthor{\bsnm{Moons},~\bfnm{Karel~GM}\binits{K.~G.}}
(\byear{2015}).
\btitle{Imputation of systematically missing predictors in an individual
  participant data meta-analysis: A generalized approach using MICE}.
\bjournal{Statistics in Medicine}
\bvolume{34}
\bpages{1841--1863}.
\end{barticle}
\endbibitem

\bibitem[\protect\citeauthoryear{Josse et~al.}{2016}]{josse2016missmda}
\begin{barticle}[author]
\bauthor{\bsnm{Josse},~\bfnm{Julie}\binits{J.}},
  \bauthor{\bsnm{Husson},~\bfnm{Fran{\c{c}}ois}\binits{F.}} \betal{et~al.}
(\byear{2016}).
\btitle{missMDA: A package for handling missing values in multivariate data
  analysis}.
\bjournal{Journal of Statistical Software}
\bvolume{70}
\bpages{1--31}.
\end{barticle}
\endbibitem

\bibitem[\protect\citeauthoryear{Kuha, Katsikatsou and
  Moustaki}{2018}]{kuha2018latent}
\begin{barticle}[author]
\bauthor{\bsnm{Kuha},~\bfnm{Jouni}\binits{J.}},
  \bauthor{\bsnm{Katsikatsou},~\bfnm{Myrsini}\binits{M.}} \AND
  \bauthor{\bsnm{Moustaki},~\bfnm{Irini}\binits{I.}}
(\byear{2018}).
\btitle{Latent variable modelling with non-ignorable item non-response:
  multigroup response propensity models for cross-national analysis}.
\bjournal{Journal of the Royal Statistical Society: Series A (Statistics in
  Society)}
\bvolume{181}
\bpages{1169--1192}.
\end{barticle}
\endbibitem

\bibitem[\protect\citeauthoryear{Manrique-Vallier and
  Reiter}{2017}]{manrique2017bayesian}
\begin{barticle}[author]
\bauthor{\bsnm{Manrique-Vallier},~\bfnm{Daniel}\binits{D.}} \AND
  \bauthor{\bsnm{Reiter},~\bfnm{Jerome~P}\binits{J.~P.}}
(\byear{2017}).
\btitle{Bayesian simultaneous edit and imputation for multivariate categorical
  data}.
\bjournal{Journal of the American Statistical Association}
\bvolume{112}
\bpages{1708--1719}.
\end{barticle}
\endbibitem

\bibitem[\protect\citeauthoryear{McLachlan and
  Peel}{2000}]{mclachlan2000finite}
\begin{bbook}[author]
\bauthor{\bsnm{McLachlan},~\bfnm{Geoffrey}\binits{G.}} \AND
  \bauthor{\bsnm{Peel},~\bfnm{David}\binits{D.}}
(\byear{2000}).
\btitle{Finite Mixture Models}.
\bpublisher{John Wiley \& Sons}, \baddress{New York}.
\end{bbook}
\endbibitem

\bibitem[\protect\citeauthoryear{Murray et~al.}{2018}]{murray2018multiple}
\begin{barticle}[author]
\bauthor{\bsnm{Murray},~\bfnm{Jared~S}\binits{J.~S.}} \betal{et~al.}
(\byear{2018}).
\btitle{Multiple imputation: A review of practical and theoretical findings}.
\bjournal{Statistical Science}
\bvolume{33}
\bpages{142--159}.
\end{barticle}
\endbibitem

\bibitem[\protect\citeauthoryear{Murray and Reiter}{2016}]{murray2016multiple}
\begin{barticle}[author]
\bauthor{\bsnm{Murray},~\bfnm{Jared~S}\binits{J.~S.}} \AND
  \bauthor{\bsnm{Reiter},~\bfnm{Jerome~P}\binits{J.~P.}}
(\byear{2016}).
\btitle{Multiple imputation of missing categorical and continuous values via
  Bayesian mixture models with local dependence}.
\bjournal{Journal of the American Statistical Association}
\bvolume{111}
\bpages{1466--1479}.
\bdoi{10.1080/01621459.2016.1174132}
\end{barticle}
\endbibitem

\bibitem[\protect\citeauthoryear{Ricci, Rokach and
  Shapira}{2015}]{ricci2015recommender}
\begin{bbook}[author]
\bauthor{\bsnm{Ricci},~\bfnm{Francesco}\binits{F.}},
  \bauthor{\bsnm{Rokach},~\bfnm{Lior}\binits{L.}} \AND
  \bauthor{\bsnm{Shapira},~\bfnm{Bracha}\binits{B.}}
(\byear{2015}).
\btitle{Recommender systems handbook}.
\bpublisher{Springer}, \baddress{New York}.
\end{bbook}
\endbibitem

\bibitem[\protect\citeauthoryear{Rubin}{1976}]{rubin1976inference}
\begin{barticle}[author]
\bauthor{\bsnm{Rubin},~\bfnm{Donald~B}\binits{D.~B.}}
(\byear{1976}).
\btitle{Inference and missing data}.
\bjournal{Biometrika}
\bvolume{63}
\bpages{581--592}.
\bdoi{10.1093/biomet/63.3.581}
\end{barticle}
\endbibitem

\bibitem[\protect\citeauthoryear{Rubin}{1987}]{rubin1987multiple}
\begin{bbook}[author]
\bauthor{\bsnm{Rubin},~\bfnm{Donald~B}\binits{D.~B.}}
(\byear{1987}).
\btitle{Multiple Imputation for Nonresponse in Surveys}.
\bpublisher{John Wiley \& Sons}, \baddress{New York}.
\end{bbook}
\endbibitem

\bibitem[\protect\citeauthoryear{Schafer}{1997}]{schafer1997analysis}
\begin{bbook}[author]
\bauthor{\bsnm{Schafer},~\bfnm{Joseph~L}\binits{J.~L.}}
(\byear{1997}).
\btitle{Analysis of Incomplete Multivariate Data}.
\bpublisher{CRC Press}, \baddress{FL}.
\end{bbook}
\endbibitem

\bibitem[\protect\citeauthoryear{Schafer and Graham}{2002}]{schafer2002missing}
\begin{barticle}[author]
\bauthor{\bsnm{Schafer},~\bfnm{Joseph~L}\binits{J.~L.}} \AND
  \bauthor{\bsnm{Graham},~\bfnm{John~W}\binits{J.~W.}}
(\byear{2002}).
\btitle{Missing data: Our view of the state of the art.}
\bjournal{Psychological methods}
\bvolume{7}
\bpages{147}.
\end{barticle}
\endbibitem

\bibitem[\protect\citeauthoryear{Si and Reiter}{2013}]{si2013nonparametric}
\begin{barticle}[author]
\bauthor{\bsnm{Si},~\bfnm{Yajuan}\binits{Y.}} \AND
  \bauthor{\bsnm{Reiter},~\bfnm{Jerome~P}\binits{J.~P.}}
(\byear{2013}).
\btitle{Nonparametric Bayesian multiple imputation for incomplete categorical
  variables in large-scale assessment surveys}.
\bjournal{Journal of Educational and Behavioral Statistics}
\bvolume{38}
\bpages{499--521}.
\bdoi{10.3102/1076998613480394}
\end{barticle}
\endbibitem

\bibitem[\protect\citeauthoryear{Teh et~al.}{2006}]{teh2006hierarchical}
\begin{barticle}[author]
\bauthor{\bsnm{Teh},~\bfnm{Yee~Whye}\binits{Y.~W.}},
  \bauthor{\bsnm{Jordan},~\bfnm{Michael~I}\binits{M.~I.}},
  \bauthor{\bsnm{Beal},~\bfnm{Matthew~J}\binits{M.~J.}} \AND
  \bauthor{\bsnm{Blei},~\bfnm{David~M}\binits{D.~M.}}
(\byear{2006}).
\btitle{Hierarchical Dirichlet processes}.
\bjournal{Journal of the American Statistical Association}
\bvolume{101}
\bpages{1566--1581}.
\bdoi{10.1198/016214506000000302}
\end{barticle}
\endbibitem

\bibitem[\protect\citeauthoryear{Van~Buuren and
  Groothuis-Oudshoorn}{2011}]{buuren2011mice}
\begin{barticle}[author]
\bauthor{\bsnm{Van~Buuren},~\bfnm{Stef}\binits{S.}} \AND
  \bauthor{\bsnm{Groothuis-Oudshoorn},~\bfnm{Karin}\binits{K.}}
(\byear{2011}).
\btitle{mice: Multivariate imputation by chained equations in R}.
\bjournal{Journal of Statistical Software}
\bvolume{45}.
\end{barticle}
\endbibitem

\bibitem[\protect\citeauthoryear{Van~Buuren and
  Oudshoorn}{1999}]{van1999flexible}
\begin{bbook}[author]
\bauthor{\bsnm{Van~Buuren},~\bfnm{Stef}\binits{S.}} \AND
  \bauthor{\bsnm{Oudshoorn},~\bfnm{Karin}\binits{K.}}
(\byear{1999}).
\btitle{Flexible Mutlivariate Imputation by MICE}.
\bpublisher{TNO}, \baddress{Leiden}.
\end{bbook}
\endbibitem

\bibitem[\protect\citeauthoryear{Van~Buuren et~al.}{2006}]{van2006fully}
\begin{barticle}[author]
\bauthor{\bsnm{Van~Buuren},~\bfnm{Stef}\binits{S.}},
  \bauthor{\bsnm{Brand},~\bfnm{Jaap~PL}\binits{J.~P.}},
  \bauthor{\bsnm{Groothuis-Oudshoorn},~\bfnm{Catharina~GM}\binits{C.~G.}} \AND
  \bauthor{\bsnm{Rubin},~\bfnm{Donald~B}\binits{D.~B.}}
(\byear{2006}).
\btitle{Fully conditional specification in multivariate imputation}.
\bjournal{Journal of Statistical Computation and Simulation}
\bvolume{76}
\bpages{1049--1064}.
\end{barticle}
\endbibitem

\bibitem[\protect\citeauthoryear{Van~der Palm, Van~der Ark and
  Vermunt}{2015}]{van2015divisive}
\begin{barticle}[author]
\bauthor{\bparticle{Van~der} \bsnm{Palm},~\bfnm{Dani{\"e}l~W}\binits{D.~W.}},
  \bauthor{\bparticle{Van~der} \bsnm{Ark},~\bfnm{L~Andries}\binits{L.~A.}} \AND
  \bauthor{\bsnm{Vermunt},~\bfnm{Jeroen~K}\binits{J.~K.}}
(\byear{2015}).
\btitle{Divisive latent class modeling as a density estimation method for
  categorical data}.
\bjournal{Journal of Classification}
\bpages{1--21}.
\bdoi{10.1007/s00357-016-9195-5}
\end{barticle}
\endbibitem

\bibitem[\protect\citeauthoryear{Vermunt et~al.}{2008}]{vermunt2008multiple}
\begin{barticle}[author]
\bauthor{\bsnm{Vermunt},~\bfnm{Jeroen~K}\binits{J.~K.}},
  \bauthor{\bsnm{Van~Ginkel},~\bfnm{Joost~R}\binits{J.~R.}},
  \bauthor{\bsnm{Der~Ark},~\bfnm{Van}\binits{V.}},
  \bauthor{\bsnm{Andries},~\bfnm{L}\binits{L.}} \AND
  \bauthor{\bsnm{Sijtsma},~\bfnm{Klaas}\binits{K.}}
(\byear{2008}).
\btitle{Multiple imputation of incomplete categorical data using latent class
  analysis}.
\bjournal{Sociological Methodology}
\bvolume{38}
\bpages{369--397}.
\bdoi{10.1111/j.1467-9531.2008.00202.x}
\end{barticle}
\endbibitem

\bibitem[\protect\citeauthoryear{Vidotto, Vermunt and
  Kaptein}{2014}]{vidotto2014multiple}
\begin{barticle}[author]
\bauthor{\bsnm{Vidotto},~\bfnm{Davide}\binits{D.}},
  \bauthor{\bsnm{Vermunt},~\bfnm{Jeroen~K}\binits{J.~K.}} \AND
  \bauthor{\bsnm{Kaptein},~\bfnm{Maurits~C}\binits{M.~C.}}
(\byear{2014}).
\btitle{Multiple imputation of missing categorical data using latent class
  models: State of art}.
\bjournal{Psychological Test and Assessment Modeling}.
\end{barticle}
\endbibitem

\bibitem[\protect\citeauthoryear{Vidotto, Vermunt and van
  Deun}{2018}]{vidotto2018bayesian}
\begin{barticle}[author]
\bauthor{\bsnm{Vidotto},~\bfnm{Davide}\binits{D.}},
  \bauthor{\bsnm{Vermunt},~\bfnm{Jeroen~K}\binits{J.~K.}} \AND
  \bauthor{\bparticle{van} \bsnm{Deun},~\bfnm{Katrijn}\binits{K.}}
(\byear{2018}).
\btitle{Bayesian multilevel latent class models for the multiple imputation of
  nested categorical data}.
\bjournal{Journal of Educational and Behavioral Statistics}
\bvolume{43}
\bpages{511--539}.
\end{barticle}
\endbibitem

\bibitem[\protect\citeauthoryear{Walker}{2007}]{walker2007sampling}
\begin{barticle}[author]
\bauthor{\bsnm{Walker},~\bfnm{Stephen~G}\binits{S.~G.}}
(\byear{2007}).
\btitle{Sampling the Dirichlet mixture model with slices}.
\bjournal{Communications in Statistics-Simulation and Computation}
\bvolume{36}
\bpages{45--54}.
\bdoi{10.1080/03610910601096262}
\end{barticle}
\endbibitem

\bibitem[\protect\citeauthoryear{White, Royston and
  Wood}{2011}]{white2011multiple}
\begin{barticle}[author]
\bauthor{\bsnm{White},~\bfnm{Ian~R}\binits{I.~R.}},
  \bauthor{\bsnm{Royston},~\bfnm{Patrick}\binits{P.}} \AND
  \bauthor{\bsnm{Wood},~\bfnm{Angela~M}\binits{A.~M.}}
(\byear{2011}).
\btitle{Multiple imputation using chained equations: Issues and guidance for
  practice}.
\bjournal{Statistics in Medicine}
\bvolume{30}
\bpages{377--399}.
\end{barticle}
\endbibitem

\bibitem[\protect\citeauthoryear{Xie and Meng}{2017}]{xie2017dissecting}
\begin{barticle}[author]
\bauthor{\bsnm{Xie},~\bfnm{Xianchao}\binits{X.}} \AND
  \bauthor{\bsnm{Meng},~\bfnm{Xiao-Li}\binits{X.-L.}}
(\byear{2017}).
\btitle{Dissecting multiple imputation from a multi-phase inference
  perspective: What happens when god's, imputer's and analyst's models are
  uncongenial?}
\bjournal{Statistica Sinica}
\bpages{1485--1545}.
\end{barticle}
\endbibitem

\end{thebibliography}

\end{document}